\documentclass[iop,apj]{emulateapj}
\usepackage{apjfonts}
\usepackage{epsf}
\usepackage{amsmath}
\usepackage{color}
\def\xi{\hbox{$X_{\rm i}$}}
\catcode`\@=11
\def\gsim{\ifmmode{\mathrel{\mathpalette\@versim>}}
    \else{$\mathrel{\mathpalette\@versim>}$}\fi}
\def\lsim{\ifmmode{\mathrel{\mathpalette\@versim<}}
    \else{$\mathrel{\mathpalette\@versim<}$}\fi}
\def\@versim#1#2{\lower 2.9truept \vbox{\baselineskip 0pt \lineskip 
    0.5truept \ialign{$\m@th#1\hfil##\hfil$\crcr#2\crcr\sim\crcr}}}
\catcode`\@=12
\def\msun{\hbox{$M_\odot$}}

\def\t9{\hbox{$t_9$}}

\def\m*{\hbox{$M_*$}}

\def\ho{\hbox{$H_\circ$}}
\def\h50{\hbox{$\ho /50$}}

\def\y1{\hbox{${\rm yr}^{-1}$}}

\begin{document}

\title{THE WFC3 GALACTIC BULGE TREASURY PROGRAM: RELATIVE AGES OF BULGE STARS OF \\ HIGH AND LOW METALLICITY*}
\altaffiltext{*}{Based on observations made with the NASA/ESA {\it Hubble
Space Telescope}, obtained at the Space Telescope Science Institute, which
is operated by the Association of Universities for Research in Astronomy, 
Inc., under NASA contract NAS 5-26555.  These observations are associated
with program GO-14759.}
\author{Alvio Renzini\altaffilmark{1},
Mario Gennaro\altaffilmark{2},
Manuela Zoccali\altaffilmark{3,4},
Thomas M. Brown\altaffilmark{2},
Jay Anderson\altaffilmark{2},
Dante Minniti\altaffilmark{4,5,6},\\
Kailash C. Sahu\altaffilmark{2},
Elena Valenti \altaffilmark{7} \&
Don A. VandenBerg\altaffilmark{8}}

\altaffiltext{1}{INAF - Osservatorio Astronomico di Padova, Vicolo dell'Osservatorio 5, 35122, Padova, Italy; \email{alvio.renzini@oapd.inaf.it}}

\altaffiltext{2}{Space Telescope Science Institute, 3700 San Martin Drive, Baltimore, MD 21218, USA}

\altaffiltext{3}{Instituto de Astrof\'{i}sica, Pontificia Universidad Cat\'{o}lica de Chile, Av. Vicu\~{n}a Mackenna 4860, Santiago , Chile}

\altaffiltext{4}{Millennium Institute of Astrophysics, Av. Vicu\~{n}a Mackenna 4860, 782-0436 Macul, Santiago, Chile}

\altaffiltext{5}{Departamento de Ciencias F'sicas, Facultad de Ciencias Exactas, Universidad Andr\'es Bello, Av. Fernandez Concha 700, Las Condes, Santiago, Chile}

\altaffiltext{6} {Vatican Observatory, V00120, Vatican City State}

\altaffiltext{7}{European Southern Observatory (ESO), Karl-Schwarzschild-strasse 2, 85748, Garching, Germany} 

\altaffiltext{8}{Department of Physics, and Astronomy, University of Victoria, P.O. Box 1700 STN CSC, Victoria, BC, V8W 2Y2, Canada}

\begin{abstract}
The HST/WFC3 multiband photometry spanning from the UV to the near-IR of four fields in the Galactic bulge, together with that for six template globular and open clusters,
are used to photometrically tag the metallicity [Fe/H] of stars in these fields after proper-motion rejecting most foreground disk contaminants. Color-magnitude diagrams and luminosity functions are then constructed, in particular for the most metal rich and most metal poor stars in each field. We do not find any significant difference between the $I$-band and $H$-band luminosity functions, hence turnoff luminosity and age, of the metal rich and metal poor components which therefore appear essentially coeval. In particular, we find that no more than $\sim 3\%$ of the metal-rich component can be $\sim 5$ Gyr old, or younger.  Conversely, theoretical luminosity functions give a good match to the observed ones for an age of $\sim 10$ Gyr.  Assuming this age is representative for the bulk of bulge stars, we then recall the observed properties of star-forming galaxies at 10 Gyr lookback time, i.e., at $z\sim 2$, and speculate about bulge formation in that context. We argue that bar formation and buckling instabilities leading to the observed boxy/peanut, X-shaped bulge may have arisen late in the history of the Milky Way galaxy, once its gas fraction had decreased compared to the high values typical of high-redshift galaxies. This paper follows the public release of the photometric and astrometric catalogs for the measured stars in the four fields.

\end{abstract}


\section{Introduction}
\label{sec:intro}
The formation of galactic bulges is one of the currently most debated issues in galaxy evolution, with efforts being concentrated in two distinct, yet complementary fronts: one at high redshift aiming to {\it see} bulges in formation, and another focusing on local galaxies, and especially on the bulge of the Milky Way, mapping its structure, dynamics  and stellar content in great detail.

High redshift observations have revealed the presence of central stellar concentration in massive galaxies, i.e., of  bulges at $z\simeq 2$, where star formation has almost completely ceased while continuing in a surrounding disk (e.g., \citealt{lang14}; \citealt{vdkk14}; \citealt{tacchella15}; \citealt{nelson16}). The disks themselves are in many respects very different from those in the nearby Universe.
They have a much higher gas fraction ($\sim 50\%$) than nearby disks \citep{tacconi10, daddi10,genzel15,scoville17}, which to first order scales as $\sim (1+z)^{2.6}$ \citep{tacconi18}, and they are more compact for a given stellar mass, with their effective radius scaling as $\sim (1+z)^{-1}$ \citep{newman12}. As a result, the surface gas density for a given stellar mass scales as $\sim (1+z)^{4.6}$, probably the most rapidly evolving galaxy property. This remains true even if adopting the slower size evolution of disks from \cite{mosleh17},  which scales as $\sim (1+z)^{-0.5}$. Moreover, high-$z$ disks are characterized by much higher gas turbulence, hence a higher velocity dispersion of stars forming out of such gas, possibly leading to {\it thick-disk} formation \citep{forster09}. Last, likely as a result of higher gas content and gas density, the star formation rate at fixed stellar mass increases as $\sim (1+z)^{2.8}$ (e.g., \citealt{ilbert15}). Of course, all of these scaling laws are affected by a sizable dispersion, typically of the order of $\sim 0.2-0.4$ dex.

In parallel with these observational findings, scenarios have been proposed and developed for the formation of central bulges in high-redshift galaxies, as due to giant clump formation and their migration and coalescence to the center \citep{immeli04,carollo07,elmegreen08,genzel08,bournaud09}, or to overall {\it violent disk instabilities} leading to the central pile up of a large amount of star-forming gas with a very short depletion time (e.g., \citealt{dekel14,tacchella16}). In both versions, rotating bulges form rapidly out of the disk, in a gas rich, highly dissipative environment, for which \cite{tadaki17} offer likely examples at $z\sim 2$ from spatially-resolved observations with ALMA.

Such formation scenarios are difficult to incorporate into the {\it pseudo-/classical-bulge} taxonomy motivated by local, low-redshift phenomenology \citep{kormendy04}, in which bulges form as a result of either dissipationless merging of sub-units (classical bulges) or of dissipationless bar formation in a gas-poor stellar disk with ensuing buckling instability of such a bar (pseudo-bulges), e.g., \cite{shen10}.  Undoubtedly in favor of this bar/buckling scenario is the fact that the Milky Way bulge is a bar and is cylindrically rotating, boxy and X/peanut-shaped (e.g., \citealt{mcwilliam10,nataf10,kunder12,zoccali14, zoccali16,ness14,ness16}), as indeed predicted by (gas-free) N-body simulations (e.g., \citealt{athana05,martinez06,shen10,gardner14}). 

Eventually, we will have to bridge the local and the high redshift evidence in order to understand how bulges have formed and acquired their present configuration.
For example, the bar/buckling instability may have developed at a relatively late time in the evolution of the Milky Way, once the gas fraction had decreased towards the present value, while a bulge was already in place and formed at very early cosmic times. Then, once formed, the bar may have captured intermediate age stars from the disk, adding them to the bulge. 
Precise age-dating of as many bulge stars as possible is then an indispensable step for reconstructing the formation history of the Galactic bulge, deriving their metallicity and age distributions, therefore quantifying the relative role of the early gas-rich formation phase, dominated by dissipation, and of the subsequent gas-poor phase, dominated by stellar dynamics. In other words, given for granted that the bulge formed from the disk, it remains to be established which fraction of the bulge stars formed early, during the gas-rich (dissipational) era of the disk, and which fraction was added later as a result of stellar dynamical (dissipationless) instabilities. 

The traditional way of measuring the age of resolved stellar systems is based on the construction of color-magnitude diagrams (CMD) and their comparison to theoretical isochrones and/or to the CMDs of systems (such as globular clusters, GC) whose age had been measured by isochrone fits.
Following this technique, \cite{ortolani95} showed that the magnitude difference between the horizontal branch clump and the main sequence turnoff (MSTO) of stars in Baade's Window (BW)
is the same as in the two most metal rich GCs of the bulge, NGC 6528 and 6553, which in turn is the same as in the inner halo GC 47 Tuc. They concluded that the Galactic bulge underwent rapid chemical enrichment to solar abundance and above, and that the bulk of the stars in the bulge ought to be older than $\sim  10$  Gyr. One limitation of this study consisted in the coarse subtraction of putative foreground disk stars, that may have eliminated some genuine intermediate age stars belonging to the bulge, if present. This limitation was overcome
by \cite{kuijken02} using CMDs for proper-motion selected members of the bulge, specifically in BW and in a Sagittarius bulge field, which again indicated a very old age. These results were based on {\it Hubble Space Telescope} (HST) optical data, hence affected by relatively high reddening, which also variable across the field.  Moving to the near-IR was the next step, to reduce the impact of differential reddening, and \cite{zoccali03} produced $JHK$ CMDs for stars in BW and compared them to a $JH$ CMD of the bulge GC NGC6528 obtained with the HST NICMOS camera. Again, this comparison showed that the bulge and clusters exhibited nearly identical magnitude differences between the horizontal branch clump and the MSTO, hence supporting an old age ($\gsim 10$ Gyr) for the bulk of the stars in the bulge.  Yet, also in this case, only a statistical foreground disk star decontamination was applied. This limitation was 
avoided by \cite{clarkson08,clarkson11} by using an HST proper-motion selected sample in the inner bulge (SWEEPS field, \citealt{sahu06}). They concluded that the best fit to the bulge population is offered by $11\pm 3$ Gyr isochrones with an upper limit of $\sim 3.2\%$ for a bulge component younger than 5 Gyr, and argued that most main sequence stars brighter than the old turnoff ought to be old blue stragglers.

All these studies were conducted on inner bulge fields near the minor axis, whereas it was still possible that a younger component could hide near the ``corners'' of the boxy bar/bulge, as one may expect for stars of disk origin. To check for this possibility, near-IR CMDs of two fields near such corners were obtained by \cite{valenti13}, finding -- after statistical field decontamination -- no appreciable age differences between such fields and those near the minor axis.

One limitation of all such studies was that the CMDs did not distinguish the metallicity of individual stars, while the bulge spans a very wide range from $\sim 1/10$ to several times the solar metallicity (e.g., \citealt{zoccali17}). In principle, the well known age-metallicity degeneracy may conjure to hide in the CMD young, metal rich stars among older, metal-poor ones. Clearly a great advantage would come from knowing the metallicity of individual stars. To this end, \cite{bensby17} have conducted a long term spectroscopic study of bulge dwarf, turnoff, and subgiant stars while being highly amplified by microlensing events. Having so far secured data for 90 objects, spectroscopy then provided metallicity, effective temperature, and gravity for each of them, hence using isochrones in the log~$T_{\rm eff}-$log $g$ plane, an age estimate for each individual star was derived. The result is that metal poor stars appear to be uniformly older than $\sim 10$ Gyr, whereas the metal rich stars appear to span a very wide range, from $\sim 2$ to $\gsim 13$ Gyr. Overall, 60\% of the stars are assigned ages younger than 10 Gyr, with a distribution peaking at 5 Gyr, and with some 25\% of stars appearing to be younger than 5 Gyr (see their Figure 16). 

Thus, this {\it spectroscopic} result appears to be at variance with the {\it photometric} ones derived from the analyses of bulge CMDs. A variety of effects may lead to such discrepant results, not least the fact that the two procedures make use of different aspects of stellar atmosphere modelling: the photometric method is using temperature-color transformations and bolometric corrections from 1D model atmospheres, whereas the spectroscopic method is using ($T_{\rm eff},g$) from spectral analysis with 3D model atmospheres. Therefore, the two methods are likely to suffer from different systematic errors, especially at high metallicity.
This discrepancy has been discussed by \cite{nataf12} and \cite{nataf16} who argued that the two methods could be reconciled by appealing to a higher helium enrichment factor
$\Delta Y/\Delta Z$ then currently adopted for the high metallicity isochrones.

To overcome  the main limitation of the photometric approach we pursued the HST {\it WFC3 Galactic Bulge Treasury Program} (GO-11664; PI T.M. Brown), where the metallicity of individual stars is estimated by a photometric 
method as described in \cite{brown09}, with some preliminary results presented in \cite{brown10} and \cite{gennaro15}. Having now also the second epoch HST observations for all our fields we are able to exploit the full dataset  secured by the project. Of course, these photometric metallicities cannot compete in accuracy with those derived from high-resolution spectroscopy, but can be measured for many thousands of stars and well below the MSTO, as opposed to $\sim 100$ objects as in the case of lensed stars. With this paper we are not trying to reconstruct the full star formation history and chemical evolution of the bulge. More modestly, we focus on a comparison between the most metal poor and most metal rich stars of the bulge with the intent of ascertaining whether there is evidence for the metal rich component being significantly younger than the metal poor one, and in particular quantify the presence, if any, of intermediate age, $\sim 5$ Gyr old stars.

\begin{figure}[] 
   \centering
   \includegraphics[width=3.5in]{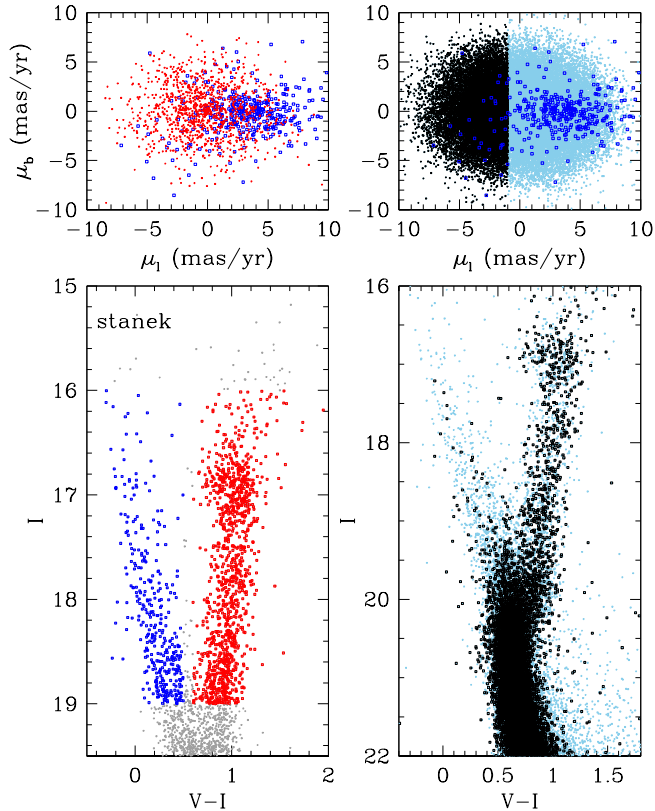} 
   \caption{This figure illustrates the procedure to clean the photometric catalog from most of the disk contamination, from the CMD of the whole sample, lower/left panel,
   to the CMD in the lower/right panel, where bona-fide bulge members are shown as black squares, and stars whose proper motions do not separate from those of disk star are shown as smaller light-blue squares. The upper panels show the vector point diagrams with the same color codes as in the lower panels. }
   \label{fig:stanek}
\end{figure}

\begin{figure*}[t]
   \centering
   \includegraphics[width=5.5in, angle=-90]{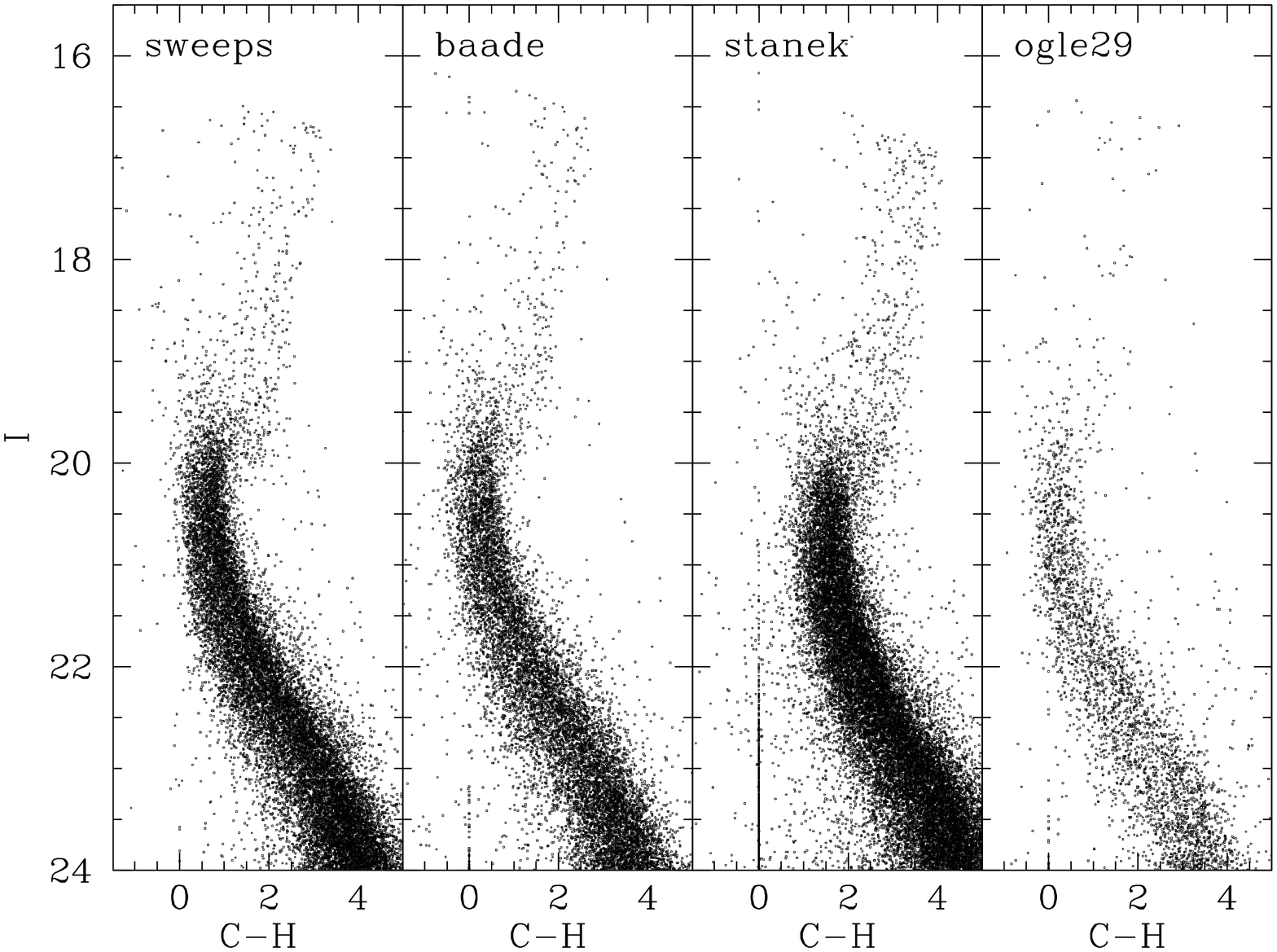} 
   \caption{The $I-(C-H)$ CMD of the four bulge fields after disk decontamination. 
}
   \label{fig:cmd}
\end{figure*}

\section{Data and bulge star selection}
\label{sec:data}
The HST/WFC3 data are those already described in Paper II for the four fields covered by the project, namely the BW, SWEEPS, Stanek and OGLE29 fields, whose main characteristics are reported in Table 1, where $R_{\rm GC, min}$ is the minimum projected distance from the Galactic center and $A_{\rm V}$ is the average extinction in each field (see Paper II for references). For the $H$-band attenuation $A_{\rm H}$ we have used the bulge reddening maps of \cite{oscar12} along with the \cite{cardelli89} infrared reddening law. Thus, the listed values of $A_{\rm V}$ and $A_{\rm H}$ have been derived in completely independent ways.

The SWEEPS field already had multiple epoch observations and therefore proper motion measurements used to produce purer samples of bulge stars \citep{clarkson08,clarkson11}. For the other three fields, 
second epoch observations were secured in 2012. Stellar photometry was performed using the code of \cite{anderson08}, ncluding updates to its current version, called two-pass "kitchen sink" (KS2). We have already released all photometry, astrometry, proper motions, and artificial star tests from the data obtained in the frame of our Treasury program.\footnote{http://dx.doi.org/10.17909/T90K6R} The observations have been performed using the WFC3 filters F390W, F555W, F814W, F110W, and F160W, hereafter referred to as $C$, $V$, $I$, $J$ and $H$, respectively, all expressed in ST-magnitudes.

\begin{table}
\centering
{
\caption{The Four Bulge Fields.}
\begin{tabular}{lcccccc}
\hline
\hline
\vspace{1 truemm}
Field     & $l$ (deg)&  $b$ (deg)   & $R_{\rm GC, min}$ (kpc) & $A_{\rm V}$ (mag) & $A_{\rm H}$ (mag)\\   
 \hline
 OGLE29 &  -6.75           &      -4.72                &  1.21 &  1.5 & 0.206\\
 Baade&        1.06            &     -3.81                &  0.58&   1.6 & 0.223\\
 SWEEPS&   1.25            &     -2.65                &  0.43&   2.0 & 0.297\\
 Stanek &       0.25           &     -2.15                 &  0.32 &  2.6 & 0.367\\
 \hline
 \hline
\vspace{-1 truemm}
\end{tabular}
}
\label{tab:fields}
\end{table}

\subsection{Selection of bulge members for age dating}
To ensure a reasonably accurate age dating of stellar populations in the bulge requires a careful selection of those stars from the global catalogs that maximize bulge membership likelihood and ensure good photometric accuracy. For each star, the catalog includes the fraction of light from contaminating neighbors, and we have selected those stars with less than 10\% contamination. We have also adopted a cut in the galactic longitude proper motion, to exclude as many disk stars as possible, as illustrated in Figure \ref{fig:stanek}  for the Stanek field. In practice, these criteria are quite similar to those used by \cite{clarkson08} and are not affected by a metallicity bias \citep{clarkson18}.

Stars painted blue in the left/lower panel, believed to be predominantly disk main sequence stars, are also shown in blue in the upper panels, showing the proper motion of these stars in the field are found predominantly at $\mu_{\rm l}\gsim 0$,  whereas the general populations shows a symmetric distribution. Stars on the red giant branch (RGB), centered on $\mu_{\rm l}=0$ by construction, are also painted red in the upper/left panel, showing that on average they have quite distinct kinematics with respect to the disk stars. We then select our {\it purified} bulge sample by picking only stars with $\mu_{\rm l}<-1$ mas/yr, as illustrated in the upper/right panel.  The galactic latitude component of the proper motion shows a symmetric distribution with similar amplitudes for both the disk and the bulge populations, and therefore it does not help discriminate between them. Of course,  disk stars are expected to have much lower velocity dispersion perpendicular to the plane, when compared to bulge stars; however, they are closer, and apparently the two factors combine to give similar dispersions in latitude proper motions. The lower/right panel shows again the CMD of the Stanek field, where the black squares show stars fulfilling this bulge-membership criterion, while the small blue squares refer to stars that do not fulfil it. The same procedure works well also for the BW, OGLE29 and the Sweeps fields, so we do not show the corresponding figures, though in OGLE 29 the proper motion separation of bulge and disk stars is not as efficient as in the other fields.

Note that a handful of stars still remains in the area occupied by the blue stars in Figure \ref{fig:stanek}.
Those are  blue stars mixed with the black ones in the upper-right panel of the figure and which extend above the MSTO in the lower/right panel. 
As shown by \cite{clarkson11}, the majority of them are blue stragglers belonging to the bulge, but some may still belong to the disk, while others may be genuine intermediate age stars belonging to the bulge. Figure \ref{fig:cmd} shows the $I-(C-H)$ CMDs 
of the four fields after the disk decontamination. This color offers the broadest baseline. 

\begin{table}
\centering
{
\caption{The six template clusters.}
\begin{tabular}{lcccccc}
\hline
\hline
Cluster    & [Fe/H] & Age (Gyr)    & DM (mag) & $E(B-V)$\\   
 \hline
 NGC 6341 &  -2.44            &      13.80                 &  14.54 & 0.022\\
 NGC 6752 &  -1.52            &      13.80                 &  12.55 & 0.050\\
 NGC   104 &  -0.80            &      12.60                 &  13.16 & 0.046\\
 NGC 5927 &  -0.47            &      12.50                 &  14.54 &  0.44\\
 NGC 6528 &  0.05           &      12.70                 &  14.48 &  0.60\\     
 NGC 6791 &  0.40             &      11.00                 &  12.84  & 0.20\\ 
\hline
\hline
\end{tabular}
}
\label{tab:clusters}
\vspace{5 truemm}
\end{table}

\begin{figure}[t] 
\vspace{-1.8 truecm}
   \centering
   \includegraphics[width=3.5in]{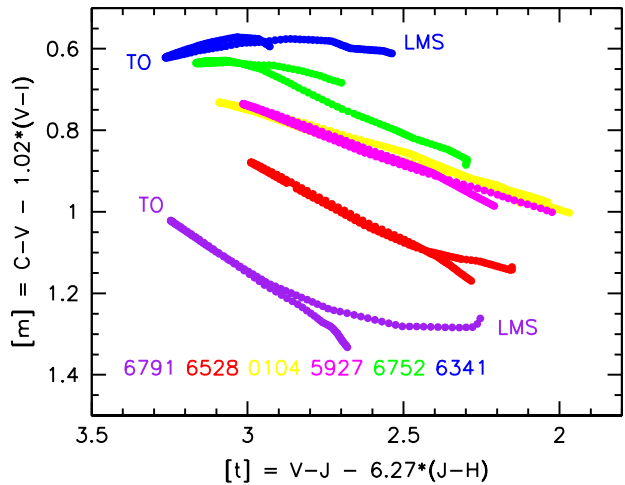} 
   \caption{The reddening-free, metallicity sensitive photometric quantity $[m]$ vs. the reddening-free, temperature-sensitive photometric quantity $[t]$ for the best-fitting isochrones of the six template clusters,  from top to bottom  NGC 6341, 6752, 5927, 104, 6528 and 6791, i.e., from the most metal poor to the most metal rich, cf. Table 2 and identified by the inserted color code. The plots include the portion of the isochrone from 3 mag below the MSTO (marked LMS) to 2 mag above it. The MSTO position is also marked in a couple of examples (marked TO) and coincides with the point of maximum $[t]$.}
   \label{fig:mt}
\end{figure}

\subsection{Cluster templates}
\label{sec:clusters}
As described in Paper I and Paper II, besides the four bulge fields, also five GCs and one old open cluster were observed with the same WFC3 filter set, in order to calibrate 
the theoretical isochrones of \cite{vandenberg14} for the STMAG system. Because the main purpose of these cluster observations was to calibrate the isochrone transformations, the observations were short and not well-dithered. This results in much noisier CMDs than one would like for detailed study of the cluster populations, but the mean locus in each CMD is sufficiently well defined to calibrate the isochrone transformations.
The cluster metallicity, age, distance modulus (DM), and reddening are reported in 
Table 2 for all six clusters, along with their ages resulting from the isochrone fits.

For each cluster the isochrones best fitting the CMDs using all five bands were obtained by marginalizing over all the parameters, such as distance modulus (DM), reddening, age, etc., plus allowing for small color shifts to perfect the fits (see Fig. 2 in \citealt{gennaro15}). These color shifts involved magnitude variations of order of a few 0.01 mag, affecting also the age-sensitive turnoff magnitudes, but implying age shifts of at most a few percent, completely negligible in the present context. This procedure gives an unusually old age for the open cluster NGC 6791, for which a 
very accurate age of 8.3$\pm 0.3$ Gyr has been derived by \cite{brogaard12}. This mismatch does not affect our dating procedures, as for this purpose we use the turnoff luminosity-age relations of the isochrones (see Section \ref{sec:lf}, which are not appreciably affected by the isochrone calibration procedures. 
The resulting best-fitting  isochrones of these clusters -- in practice just a smooth representation of the CMD ridge lines -- are then used to map the $[m]-[t]$ plane in terms of metallicity, where the reddening-free metallicity $[m]$ and temperature $[t] $ indices are defined as (cf.\ Paper I and II):

\begin{equation}
[m] = C-V -1.02\times (V-I)
\end{equation}
and
\begin{equation}
[t]= V-J-6.27\times (J-H).
\end{equation}
These indices are reddening free for the adopted reddening law from \cite{fitzpatrick99} with $R_{\rm V}=2.5$ \citep{nataf13}, though using a different value (e.g., $R_{\rm V}=3.1$) would not affect any of our conclusions.
The coefficients in these two equations are slightly different  from those adopted in Papers I and II because in those papers we used the Vega magnitude system, whereas we now use 
the STMAG system. Here we use the cluster best-fitting isochrones to map the $[m][t]$ plane in terms of metallicity: Figure \ref{fig:mt} shows these cluster loci in the $[m]-[t]$ plane, having selected the portion of the isochrone extending from 
$\sim 3$ magnitudes below the MSTO to $\sim 2$ magnitudes above it, thus including the whole subgiant branch (SGB) and the lower part of the RGB. This will also be the part of the bulge CMDs that will be used for age-dating purposes. We first notice that the six clusters separate in this plot according to their metallicity, increasing from [Fe/H]=$-2.44$ (NGC~6341) to +0.40 (NGC~6791)\footnote{For convenience these cluster  best fitting isochrones can be found in www.astro.puc.cl/$\sim$mzoccali/OnlineData}. 

The bulge cluster NGC 5927 appears to overlap with NGC 104, as being just slightly more metal rich, in spite of a nominal 0.33 dex difference reported in Table 2
from the most recent study of this cluster \citep{mura18}. We have no explanation for this apparent mismatch but note that there may be a slight difference in the reddening law in the direction of the bulge and in the direction of NGC 104, such that this cluster and NGC 5927 could overlap in the $[m]-[t]$ plot even if their metallicities differ by a factor of $\sim 2$.

 In the $[m]-[t]$ plot the lower MS starts were marked LMS on Figure \ref{fig:mt}, then the MS proceeds towards higher values of the temperature-sensitive parameter $[t]$ and the MSTO (marked TO) corresponds to the maximum value of $[t]$. The SGB then traces down to almost perfectly overlap with the MS locus until reaching the RGB. As anticipated in Paper I, upper RGB stars would follow a different metallicity-$[m]-[t]$ relation, but in this paper we use only the lower part of the RGB, which basically overlaps with the MS in the $[m][t]$ plot. Thus,  Figure \ref{fig:mt} shows that the MS, SGB, and lower RGB stars follow the same relation and therefore these sequences are excellent constant-metallicity loci, when excluding brighter giants.
The MS+SGB+RGB locus of NGC 6528 ([Fe/H]$\simeq 0$) can be taken as the dividing line between sub-solar and super-solar metallicities. 
 
 We can also notice that a line joining the highest $[t]$ points of all the cluster loci correspond to the cluster MSTOs, hence stars falling on the left side of the line  should be younger than the clusters. However, the photometry is affected by errors, and the coefficient of 6.27 in Equation (2) works as an amplifier, such that even a small error in $J-H$ becomes a sizeable error in $[t]$.  Therefore, we use this plot only to infer metallicities, not ages.
 
\section{Separating metal rich and metal poor stars for the four fields}
\label{sec:4cmd}
In this section, we first plot the MS+SGB+RGB stars in the four fields on the $[m]-[t]$ plane,  in order to photometrically separate them according to their metallicity, and then present the resulting CMDs color-coded according to the derived stellar metallicities.

 Metallicity distributions for the four fields derived from the $[m][t]$ indices were presented in Paper II and show reasonable agreement with those obtained spectroscopically. More specifically,
 the metallicity range of the various fields, median, and the gradient with the distance from the Galactic center are consistent with those derived by the GIBS survey \citep{zoccali17}.
 However, the distributions of photometric metallicities shown in Paper II are all unimodal, while those from spectroscopy are markedly bimodal for the fields with similar distance from the Galactic center, showing  a minimum at [Fe/H]$\simeq 0$. Most likely, the larger errors affecting the photometric metallicities have the effect of washing out this bimodality. 

\begin{figure}[t] 
   \centering
   \includegraphics[width=3.3in]{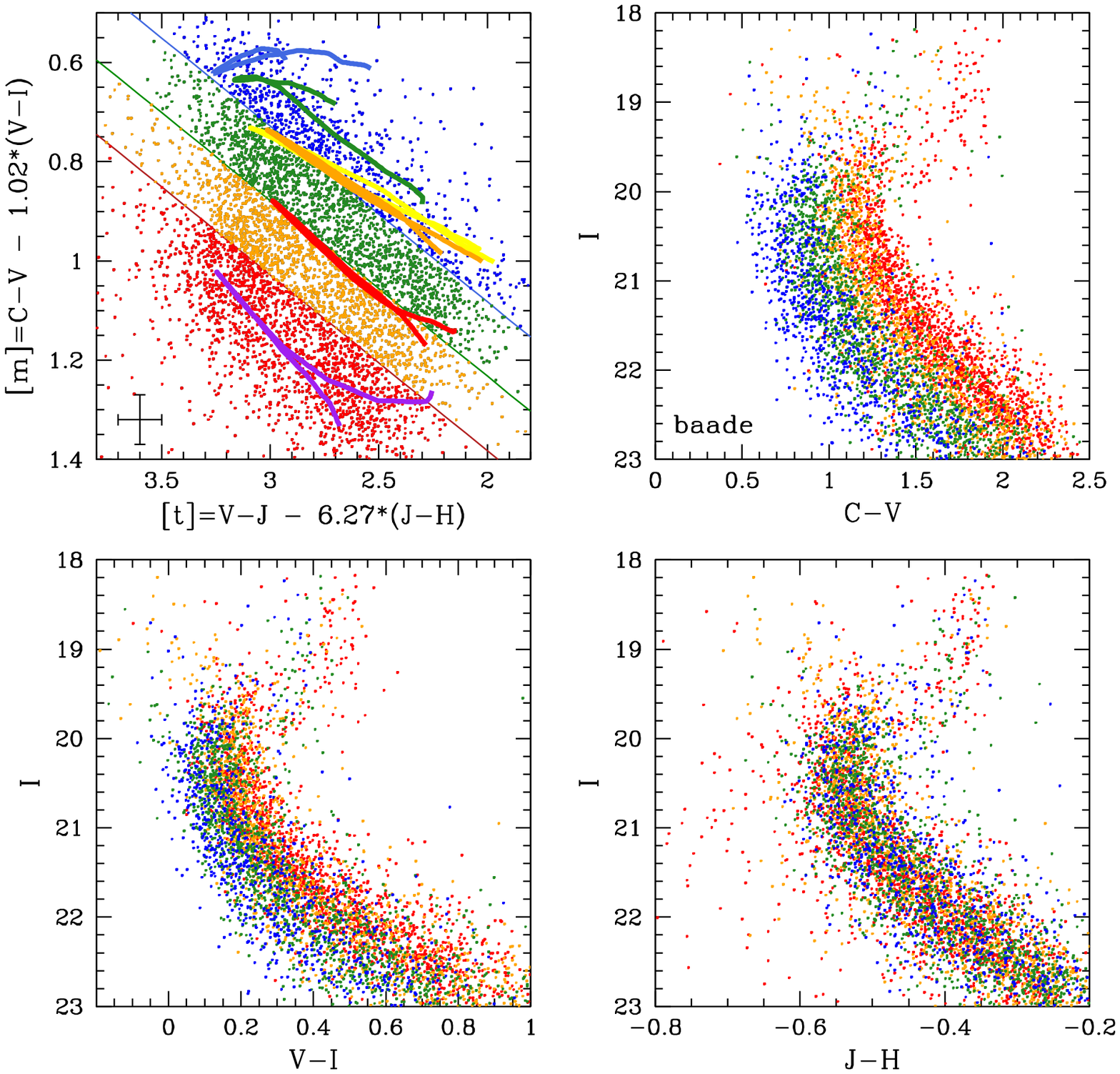} 
    \includegraphics[width=3.3in]{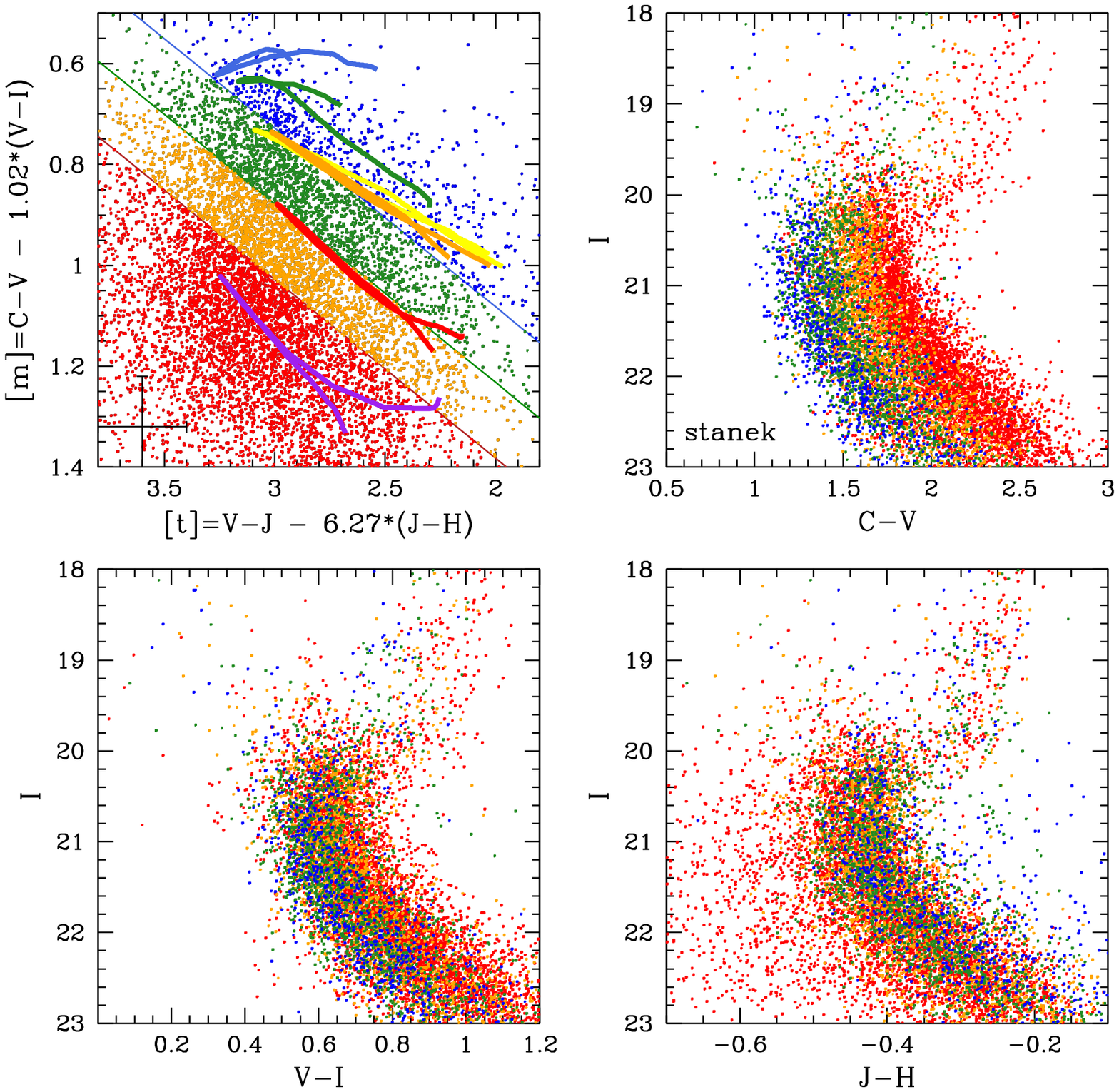} 
   \caption{Metallicity tagging (upper/left panel of each quartet) and various CMDs for bulge proper-motion selected stars in Baade's Window (top panels) and in the Stanek field (bottom panels). In the metallicity-tagging panels, the cluster template loci from Figure \ref{fig:mt} are reproduced. Stars in the CMDs are colored according to the metallicity tagging from the corresponding upper/left panel. Maximum allowed errors in 
   $[m]$ and $[t]$ are indicated in the upper/left panel.  }
   \label{fig:baade4}
\end{figure}

\begin{figure}[t] 
   \centering
   \includegraphics[width=3.3in]{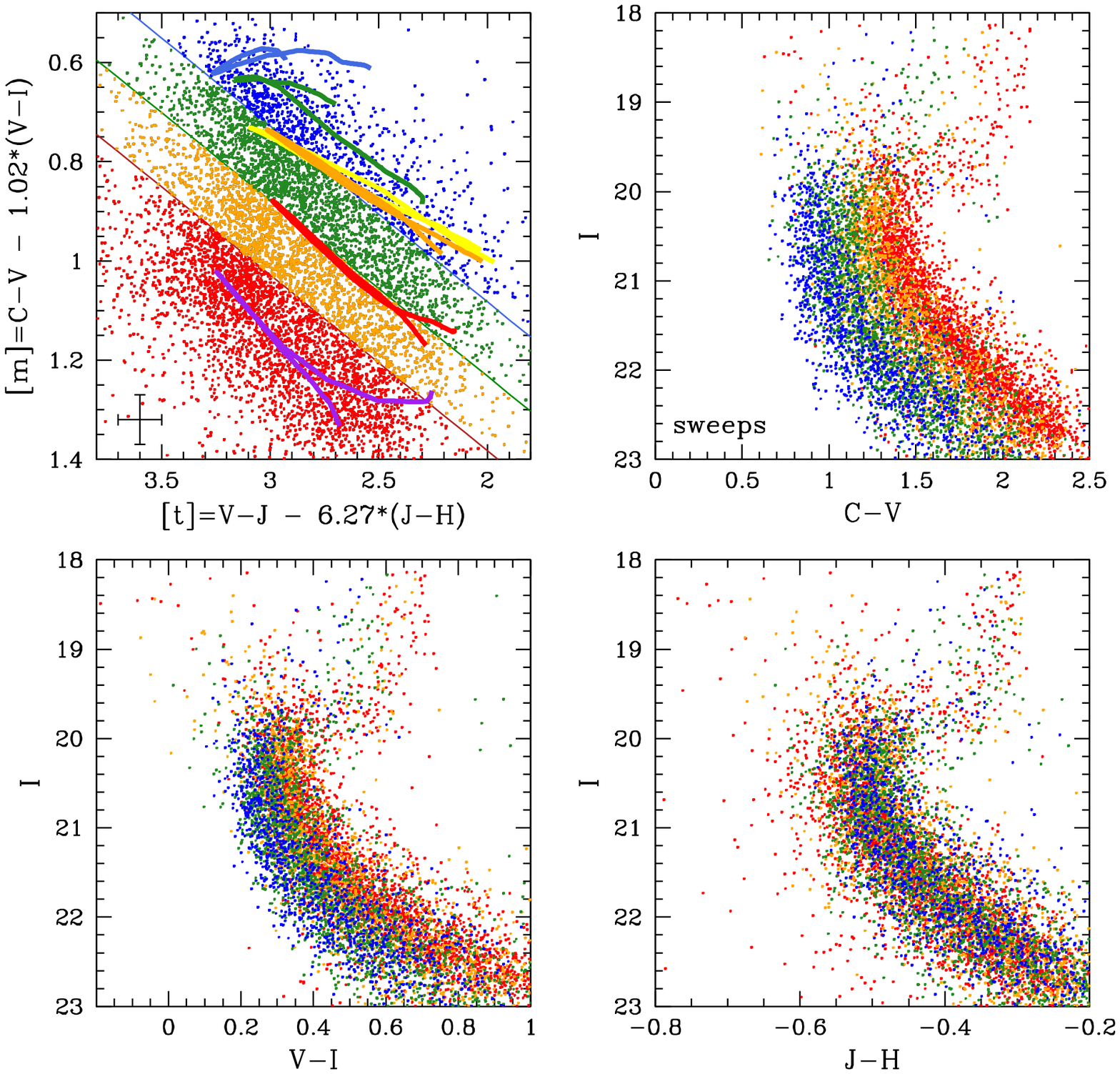} 
   \ \includegraphics[width=3.3in]{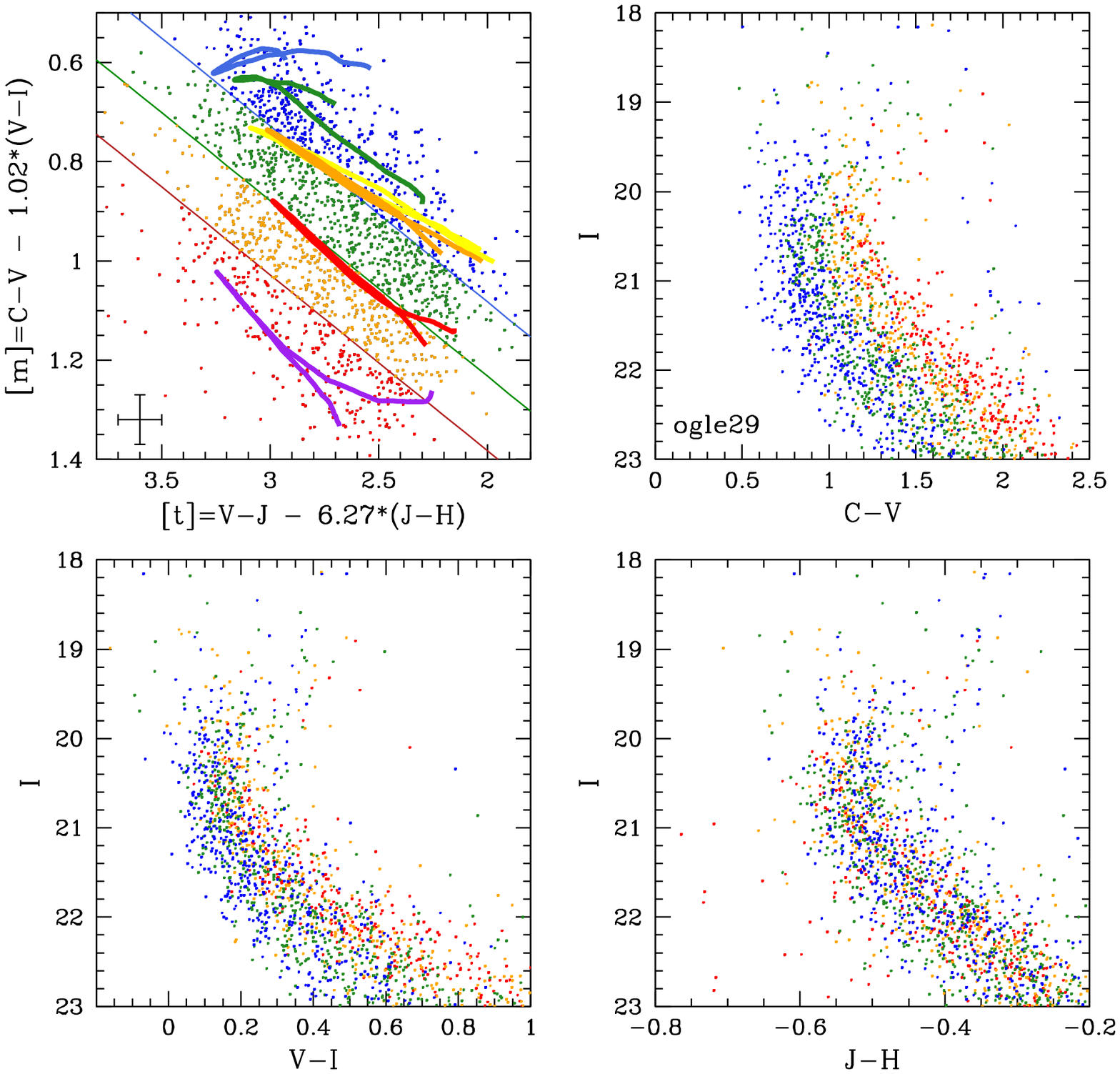} 
   \caption{The same as Figure \ref{fig:baade4} but for the SWEEPS and OGLE29 fields. For this latter field, notice the reduced metal rich/poor ratio, relative to the other fields, already mentioned in Paper II.}
   \label{fig:sweeps4}
\end{figure}

The metallicity binning and three representative CMDs for each of the four fields are shown in Figures \ref{fig:baade4} and \ref{fig:sweeps4}. The upper/left panel of each figure shows the metallicity assignments in the $[m][t]$ plane, having the template cluster loci as metallicity standards. Thus blue corresponds to stars more metal poor than NGC104 and NGC 5927, or [Fe/H]$\lsim -0.7$, green corresponds to $-0.7\lsim\rm [Fe/H]\lsim 0.0$, orange to $0.0\lsim\rm [Fe/H]\lsim +0.2$ and red to
Fe/H]$\gsim +0.2$. The error bars in each upper/left panel represents the {\it maximum} error in $[m]$ and $[t]$ and stars with larger errors in either $[m]$ or $[t]$ are not plotted in this as well as in the other panels of these figures.

 
With one exception, the CMDs of the various fields split according to metallicity, with the most metal rich stars on the red side of the diagrams and the most metal poor ones on the blue side, as expected. The exception is seen in the infrared CMD, where stars of different metallicities do not split in the $J-H$ color. The color split is maximum for colors involving the $C$ magnitude, as expected since this UV passband is the one most sensitive to metal-line blanketing (Paper I). For the rest, the figures speak for themselves, but notice that the width of the sequences is almost identical in the various fields, notwithstanding differences in crowding. 
Having for each star the photometric errors, we have evaluated the errors affecting $[m]$ and $[t]$, and only stars with $[m]$ and $[t]$ errors less than 0.05 and 0.1 mag, respectively, are plotted for the BW, OGLE29, and SWEEPS fields, and errors less than 0.2 mag for the Stanek field.  

\begin{figure}
 \vspace{-7 truemm}
   \centering
    \includegraphics[width=5in]{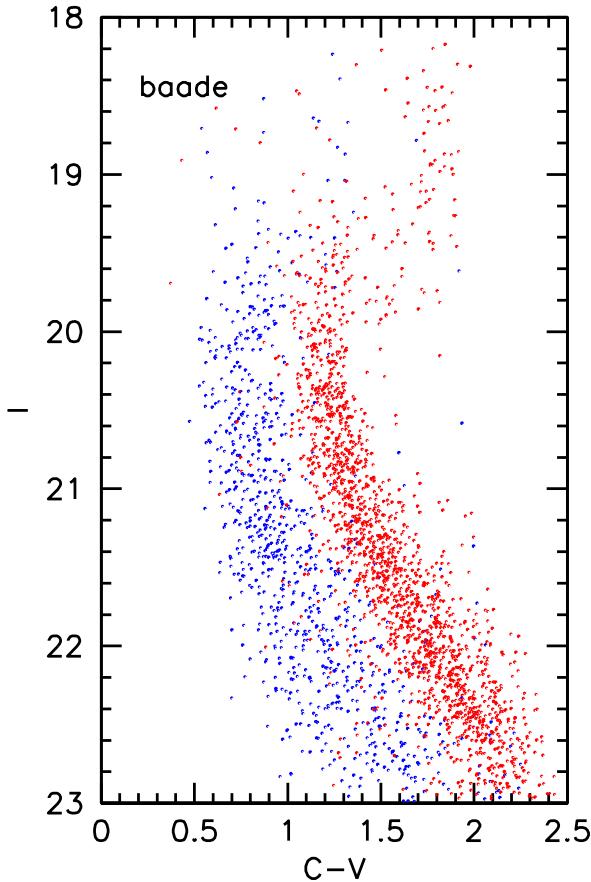} 
   \caption{The CMD for the most metal poor and most metal rich bins of BW stars, as in Fig. 4 (blue and red stars, respectively).}
   \label{fig:baaderp}
\end{figure}

As emphasized above, this photometric metallicity tagging is imperfect, as errors alone generate a spurious spread in the assigned metallicities, so we must expect a great deal of migration across the formal metallicity boundaries shown in Figures \ref{fig:baade4} and \ref{fig:sweeps4}. In order to minimize this effect, age estimates will be restricted to a comparison of the most metal poor and most metal rich bins, shown as blue and red points in Figures \ref{fig:baade4} and \ref{fig:sweeps4}. Figure \ref{fig:baaderp} shows the CMD for the two components in the BW field. In practice, the metal-rich component includes stars with [Fe/H]$\gsim 0.2$, and the metal-poor components stars with [Fe/H]$\lsim -0.7$, thus leaving a metallicity gap of almost one dex between them.

\section{Observed and Synthetic Luminosity Functions}
\label{sec:lf}
Following the early proposal by \cite{Paczynski84}, we use the stellar luminosity function (LF) as an age diagnostic. 
In this section we present the observed $I$-band and $H$-band LFs of the metal-poor and metal-rich components for our fields and compare them to synthetic LFs with various ages and metallicities. 

For the construction of the synthetic LFs, we simulated simple stellar populations (SSP) using stellar models and artificial star tests.
Artificial stars are considered detected when they satisfy the same selection criteria as the real stars used to build the chemically-tagged observed populations.
In detail, artificial stars had to be detected in all 5 bands, and have maximum errors that match the maximum allowed errors used the to build the $[m] $and $[t]$ coefficients for the real stars. The slope of the initial mass function
(IMF) is fixed to $s = -2.3$ (i.e., Salpeter IMF) as the
mass range we are dealing with is above the observed flattening of the bulge IMF \citep{zoccali00}. Moreover, for
each individual field, the distance distribution along the line
of sight is taken from \cite{wegg13}, and an extinction (Table 1) is applied. We include a
binary fraction of 30\%, with a uniform mass ratio distribution
between 0 and 1. Finally, \cite{vandenberg14} isochrones are used adopting a helium enrichment $\Delta Y/\Delta Z = 1.5$.
The $\alpha$-element enhancement is assumed to be [$\alpha$/Fe]= 0.4 up to [Fe/H]=$-0.6$, then declining linearly with increasing [Fe/H], reaching 0 at [Fe/H ]= 0.2
(see e.g., \citealt{lecureur07,johnson11,rich12,bensby17}).  Finally, MARCS
model atmospheres and synthetic spectra \citep{gustafsson08} are used to generate the intrinsic stellar
magnitudes in all 5 bands. The latter are transformed into noisy, incomplete measurements
using the results of the artificial star tests, with the selection criteria described above in this section. 
Each simulated SSP and relative LF have been constructed for 100,000 stars, with over 60,000 brighter than $I=22$. 

Above the $I=22$ limit, the number of metal rich stars used for the observed LFs is  4177,  2437, 1712 and 185, respectively for the Stanek, SWEEPS, BW, and OGLE29 fields, reflecting the surface brightness of the four fields and a tighter proper motion cut for the OGLE29 field.  As a result, the statistics for the OGLE29 field are quite poor, and and we exclude this field from further analysis.
The observed LFs include all stars detected in all five bands, regardless of their photometric errors, because very few stars fail to meet the above error criteria, given that we only make use of stars brighter than $I\sim 22$ for which artificial star tests indicate a completeness well above 80\%. 
 
\begin{figure}[t]
   \centering
    \includegraphics[width=3.9in]{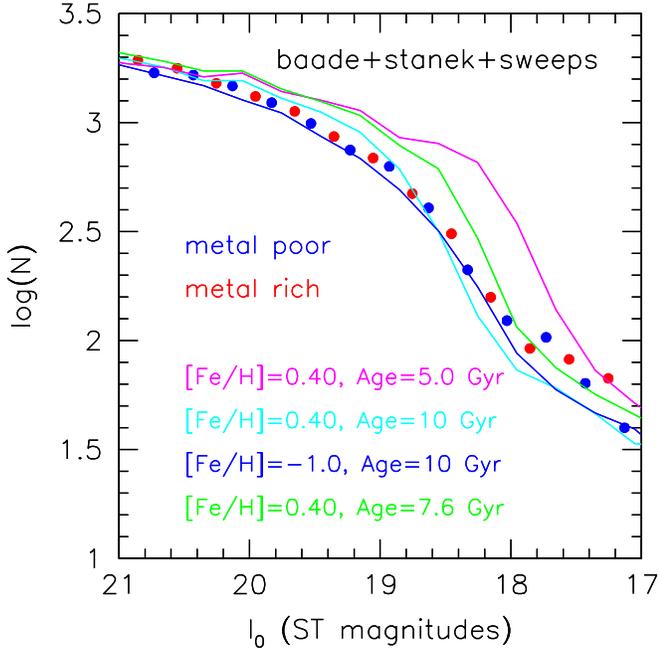} 
   \caption{The coadded luminosity functions for the most metal poor and most metal rich bins (blue and red stars, respectively) of stars as in Figure \ref{fig:baaderp} for the BW and similar selections for the Stanek and SWEEPS fields. All fields have been corrected for extinction. The simulated LFs for the metal rich component and two different ages and for the metal poor component are also shown. The number counts in the vertical axes refer to the metal rich component, whereas all of the other luminosity functions have been normalised to have the same number of stars near $I_\circ =20.5$. In particular, the number counts for the metal poor component are scaled up a factor $\sim 4$ with respect to the metal rich component.}
   \label{fig:lf3fields}
\end{figure}

In order to achieve a statistically robust result, we then proceed to coadd the LFs of the BW, Stanek and SWEEPS fields. To this end, the LFs of the Stanek and SWEEPS fields are first shifted to the same extinction of the BW field, with $A_{\rm I}=0.56A_{\rm V}$, as appropriate for the adopted reddening law (see Section \ref{sec:clusters}), with $A_{\rm V}$ values from Table \ref{tab:fields}, and then coadded. Finally, the extinction correction for the BW field is applied to both the observed, coadded LFs and the synthetic LFs. No adjustment for distance differences was necessary. The result is shown in Figure~\ref{fig:lf3fields}.

The figure shows that the LFs of the bulge metal rich and metal poor components are quite similar to each other and close to the synthetic LFs for an age of $\sim 10$ Gyr. The synthetic LF for [Fe/H]=0.4 and an age of 5 Gyr largely overpredicts the number of stars in the most age-sensitive range $17.8<I_\circ<19.0$~mag.
In this range, the three fields together include 1719 metal rich stars, whereas the normalised, synthetic LF for an age of 5 Gyr includes 2902 stars, 1183 stars more than the observed LF.  The standard deviation of the stellar counts in the mentioned $I$-band range being $\sqrt{1719}\sim 40$  stars, the number expected if the age of the bulge metal rich population was 5 Gyr is $\sim 30\sigma$ away from the observed number. With the same elementary statistics, we can infer that no more than $\sim 40/1183$,  i.e., $\sim 3\%$ of the metal rich stars in the studied fields can be $\sim 5$ Gyr old, or younger. Only above $I_\circ\sim 18$ does there appear to be a modest excess in both the metal rich and metal poor components, relative to the corresponding 10 Gyr synthetic LF. This excess is consistent with being entirely due to blue stragglers \citep{clarkson11}, though we cannot rule out that it may include some genuine intermediate age stars.
We emphasize that our focus is primarily on the {\it relative} ages of the most metal poor and most metal rich components of the bulge, not on their absolute ages, though $\sim 10$ Gyr appears to be a fair estimate for both.

\begin{figure}[t]
   \centering
    \includegraphics[width=3.9in]{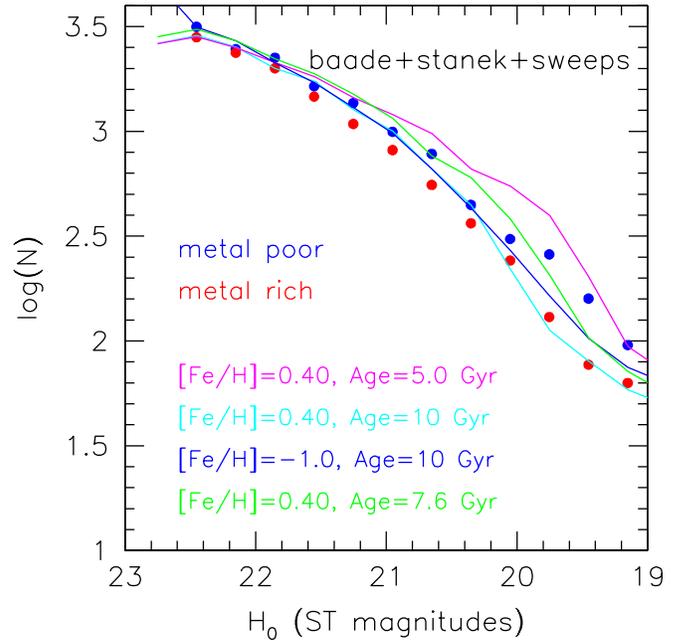} 
   \caption{The same as Figure \ref{fig:lf3fields} but for the $H$ band.  All luminosity functions have been normalised to have the same number of stars near $H_\circ =22$~mag.}
   \label{fig:hband}
\end{figure}

We emphasize that a critical entry in this procedure is the extinction correction corresponding to the $A_{\rm V}$ values of the various fields, as reported in Table 1 from Paper II, which are used to bring the luminosity functions of the three fields to a common extinction in the $I$ band before coadding them, and then correct for it before comparing to the simulated luminosity functions. On the other hand, the precise shape of the bulge reddening law remains uncertain (see e.g., \citealt{sumi04,nataf13,nataf16,majaess16,alonso17}) and so are the $A_{\rm I}$ values adopted for our fields. One way to minimize the effect of these uncertainties on the age dating is to work in the near-IR, and our HST/WFC3 database has the advantage of including also the $H$ band. Thus, we have repeated the same procedure for the $H$ band, deriving the $A_{\rm H}$ values of the various fields from the bulge reddening maps of \cite{oscar12} and adopting the reddening law of \cite{cardelli89}.
Adopting different laws, such as \cite{fitzpatrick99} or \cite{nishiyama09}, would give differences of order of $\sim 0.1$ mag in $A_{\rm H}$, corresponding to age differences of $\sim 10\%$. 

Figure \ref{fig:hband}, analogous to Figure \ref{fig:lf3fields}, shows that also in the $H$ band the LFs of the bulge metal rich and metal poor components are quite similar to each other and close to the synthetic LFs for an age of $\sim 10$ Gyr. In the age sensitive range $19.45<H<20.65$~mag, the three fields together include 1368 metal rich stars, whereas the normalised, synthetic LF for an age of 5 Gyr and [Fe/H]=0.40 includes 2781 stars, 1413 stars more than the observed LF. These numbers are quite similar to those we have derived for the $I$-band LFs, and indeed they confirm that no more than $\sim 3\%$ of the metal rich stars in the studied fields can be $\sim 5$ Gyr old, or younger.  The figures include also 7.6 Gyr, metal rich, synthetic LFs. If the majority of metal rich stars were to be younger than 7 Gyr the red points in these figures should lie above and to the right of the green lines.

Finally, as a sanity check, we derived extinctions in all five bands and in all four fields using our HST data themselves. To this end we have forced the isochcrones 
to match the slope of the main sequence and the luminosity of the {\it kink}, i.e., the turn to bluer infrared colors on the lower main sequence, which is due to the formation in the stellar envelope of the H$_2$ molecule (e.g., see \citealt{correnti16} and references therein). The resulting $A_{\rm I}$ and $A_{\rm H}$ values are within a few 0.01 magnitudes from those used in this paper and taken from the literature, giving us further confidence that our conclusions are not appreciably affected by the choice of the adopted extinctions.

\section{Discussion and Conclusions}

This approach to age-dating the bulge stellar populations is deliberately kept as simple as possible, with all assumptions clearly spelled out. This is meant to ensure that our results are easily {\it reproduced} (or {\it falsified}), because our reduced HST data and the relative  photometric and astrometric catalogs have already been publicly released. Thus, other teams may test our approach, perhaps adopting different assumptions concerning reddening, distance distributions along the various lines of sight, and/or using different sets of isochrones. In \cite{gennaro15}
we adopted a more sophisticated methodology, based on extensive simulations in an advanced Bayesian framework. While the two results consistently concur 
in indicating a dominantly old age for bulge stars of all metallicities, the simulations/Bayesian approach would require a much harder effort from other teams to 
reproduce/falsify it. 

Having photometrically separated bulge stars into metal-rich ([Fe/H]$\gsim +0.2$) and metal poor ([Fe/H]$\lsim -0.7$) components, we find that their ($I$- and $H$-band) luminosity functions are fairly similar to each other in all four fields. In particular, this is the case for the coadded LFs of the three well populated fields, which are both well matched by theoretical luminosity functions for an age of $\sim10$ Gyr.  There appears to be no need to invoke an intermediate age component, say $\sim 5$ Gyr old,  in order to match the observed luminosity functions, most notably for the highest metallicity component.  This is true for both the $I$-band and the $H$-band LFs, the latter one being far less prone to uncertainties in reddening and reddening law.

Thus, there is a tension between the present analysis and 
the result of \cite{bensby17}, in which $\sim 60\%$ of the bulge stars are found to be younger than 10 Gyr and $\sim 25\%$ younger than 5 Gyr, a fraction that rises to 40\% when considering only stars with super-solar metallicity, with $\sim 70\%$ younger than 7 Gyr. The present analysis, using
proper-motion cleaned bulge samples in combination with photometric metallicities, basically confirms the previous results from the {\it photometric} age dating method, such as those quoted in the introduction. In particular, we estimate that such a young population should not exceed $\sim 3\%$ of the metal rich component, in agreement with the result of \cite{clarkson11}, who gave an upper limit of 3.4\% (of the total population) under the most conservative scenario for the blue stragglers population of the bulge, which we have not considered. Of course, the metal rich stars ought to be somewhat younger than the metal poor ones, also because of their lower $\alpha$-element enhancement, but these data do not allow us to firmly discriminate age differences of the order of $\sim 1$ Gyr.

Our result is also at variance with a recent one based on a fraction of the same HST data.  \cite{bernard18} have used part of the same photometric and astrometric catalogs to conclude that 11\% of the bulge stars are younger than 5 Gyr. Bernard et al. used only the $V$- and $I$-band data, and hence did not distinguish stars of different metallicities, constructed synthetic CMDs using stellar population templates of various ages and metallicities, and then simulated all possible star formation histories. The weight of each individual sub-population was then determined by a Poissonian equivalent to the $\chi^2$ statistics and therefore derived a full star formation history for the bulge and associated chemical evolution. Their result is in qualitative agreement with that of \cite{bensby17}.

We are not in the position to check or falsify these results. Concerning the spectroscopic approach of \cite{bensby17}, it would be interesting to know the (unlensed) photometry of the 90 stars in their sample, in order to compare their positions in the CMDs of representative bulge fields for which the {\it photometric} method gives uniformly old ages. We note, however, that the two results agree in concluding that the metal poor component is uniformly old. Now, because the LF of the metal rich component is so similar to that of the metal poor one, the inference is that their ages must also be similar, {\it independent of any assumption concerning reddening law and attenuation of the various fields.} Indeed, this comparison is completely independent of the adopted reddening and distance distributions, being a differential comparison of LFs in the same fields. So, the discrepancy with \cite{bensby17} is confined to the high metallicity population. One may suspect that 
the {\it spectroscopic} method could be affected by larger uncertainties in the model atmosphere analysis for stars of super-solar
metallicities, as such spectra are more complex than those in the metal poor regime.

\begin{figure}[t] 
   \centering
   \vspace{-17 truemm}
    \includegraphics[width=3.5in]{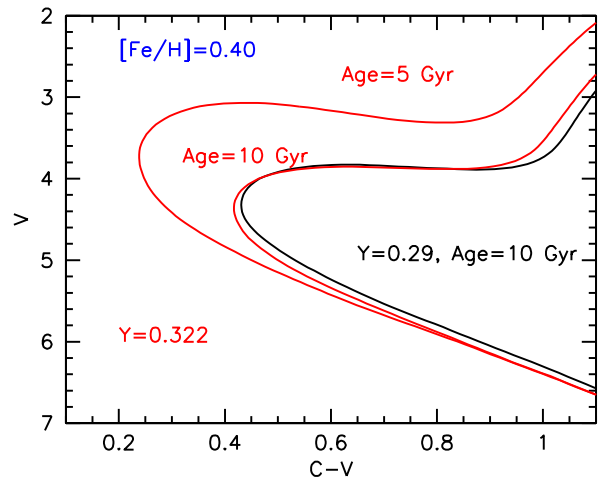} 
\vspace{-8 truemm}
   \caption{The high metallicity [Fe/H]=0.40 isochrones of VandenBerg et al. (2014) for the indicated ages, and color-coded for two values of the helium content. }
   \label{fig:helium}
\end{figure}

\cite{nataf12} have argued that the ages from the  photometric and spectroscopic methods could be reconciled if one appeals to a very high helium enrichment in the bulge, with $\Delta Y/\Delta Z\simeq 5$, as opposed to the canonical 1.5 adopted here and in previous works.
We notice that the theoretical isochrones provide  luminosity functions that for a given age are fairly insensitive to metallicity, at least for the adopted helium enrichment parameter $\Delta Y/\Delta Z$. Could a younger metal rich component disguise itself as old by having a substantially higher helium?  
For high metallicity, the \cite{vandenberg14} isochrones reach only up to $Y=0.322$  and the corresponding effect is illustrated in Figure \ref{fig:helium}.
The effect appears to be much smaller than needed to have a 5 Gyr old, metal and helium rich population disguise itself as an old population with normal
helium. Yet, if the helium enrichment was as high as $\Delta Y/\Delta Z\simeq 5$, the most metal rich stars in the bulge ([Fe/H]$\simeq 0.40$, $Z\simeq0.05$) would have $Y\simeq 0.50$, a really extreme value for which we have no independent evidence. 

Concerning the \cite{bernard18} study, it would be important to know which are the individual stars that drive their result to the presence of an intermediate age component. At the moment, they are hidden in the $\chi^2$-like algorithm, which, of course, tends to make use of all allowed degrees of freedom to perfect the fit to the data. Because both our study and their study are based on the same data, the case would be immediately evident once the CMD locations of such putative intermediate age stars were known and shown in a plot like Figure \ref{fig:baaderp}.

So, if indeed the bulk of stars in the MW bulge are $\sim 10$ Gyr old, it is at a 10 Gyr lookback time, viz at $z\sim 2$, that we can look for galaxies in the act of brewing their bulges, as already noted in the literature (e.g., \citealt{zoccali14, gonzalez16,nataf17}). Thus, with the bulge mass of $\sim 2\times 10^{10}\,\msun$ (\citealt{portail15,valenti16}), a star-forming galaxy at $z=2$ of such mass would typically have a half-mass radius of $\sim 1.5$ kpc (see, e.g.,  Figure 10 in \citealt{mosleh17}), which indeed roughly corresponds to the half-mass radius of the MW bulge. A $z=2$ star-forming galaxy of this mass is typically a compact, rotating, gas-rich disk with rotation velocity to
velocity-dispersion ratio $v/\sigma\simeq 2-6$ (\citealt{forster09,forster18}). We then argue that the age dating presented here favors a scenario in which the bulk of bulge stars have formed in the inner part of an early, rotating, actively star-forming disk which, furthermore, may have been subject to instabilities resulting in radial gas inflow and promoting enhanced star formation (e.g., \citealt{bournaud16,tacchella16}). 

In the following 10 Gyr, the gas fraction secularly dropped by a factor $\sim 10$, the disk size grew by more than a factor of $\sim 3$ as a result of the effective radius $(1+z)^{-1}$ scaling at fixed mass, plus the effect of a factor of $\sim 3$ increase in total stellar mass. Then the disk grew more and more stellar dominated, hence 
increasingly prone to bar-formation and ensuing buckling instabilities, that finally gave to the bulge its present shape. Indeed, bar formation in disk galaxies seems to be a relatively late time event, as 
the fraction of barred galaxies appears to drops very rapidly with increasing redshift \citep{sheth08,melvin14}. The presence of a bar in our galaxy has likely resulted in disk 
stars near the end of the bar to be swallowed by it and incorporated into the bulge. So, why do we miss to  see a major intermediate age component in our fields?
One possibility is that most of the stars in the inner disk, $\sim 2-3$ kpc from the center, also formed very early, when the Milky Way was substantially smaller and its specific star formation rate very high. Thus, most bar-captured stars may be very similar in age and composition to the stars that formed nearer to the center, at the main epoch of bulge formation. Precise age dating of stars in the inner disk, say, near the end of the bar, should soon become possible with Gaia, hence testing this conjecture.

In this scenario, another issue that remains to be understood is 
how the MW bulge ceased almost completely to form stars a long time ago, i.e., got {\it quenched}, and managed to remain quenched ever since.
In the simulations of e.g., \cite{tacchella16}, bulge quenching results from the rapid consumption of all the gas by star formation and polar outflows following a {\it compaction} event.  But, if so, why  did the bulge not resume the formation of stars while the disk was still doing so and growing in size? It is generally believed that gas inflow is necessary to keep stars forming in galaxies, given their short gas-depletion timescale (e.g., \citealt{tacconi18}, and references therein). Now, if disks have to grow in size following the $(1+z)^{-1}$ scaling, hence with increasing angular momentum, then gas accretion has to take place in an ordered fashion and preferentially co-rotating with the disk itself. This suggests that such accretion flows come in predominantly through the equatorial plane of galaxies, adding gas -- and then stars -- to their outer rim, with higher and higher angular momentum as time goes on. Thus, one possibility to keep a bulge quenched is that the gas accreted via streams at late times comes in with too high angular momentum to be able to reach down to the bulge, which then would remain starved and almost completely passively evolving.

Yet, these speculations are based on having taken for granted that the bulge is dominated by $\sim 10$ Gyr old stars. However, the tension between the present photometric ages and the microlensing/spectroscopic ones remains and we are not able to resolve it here. It would help to compare our CMDs with the actual CMD for the \cite{bensby17} unlensed stars that have experienced a microlensing event. In any case, our photometric and astrometric catalogs of the four fields are publicly available, thus offering to the community the opportunity to check our procedures and conduct an independent analysis of the data.

\section*{Acknowledgments}
Support for HST Program 11664 was provided by NASA through a grant from STScI, which is operated by AURA, Inc., under NASA contract NAS 5-26555. AR is grateful to the Instituto de Astrofisica of the Pontificia Universitad Catolica and the Millennium Institute of Astrophysics (Santiago, Chile) for their kind support and hospitality when this paper was first drafted; partial support was also provided by a INAF/PRIN-2018 grant (PI Leslie Hunt). MZ and DM acknowledge support by the Ministry for the Economy, Development,
and Tourism's Programa Iniciativa  Cient\'{i}fica Milenio through grant  IC120009, awarded to
Millenium Institute of Astrophysics (MAS), by  the  BASAL  CATA  Center  for Astrophysics and Associated Technologies
through grant PFB-06, and by FONDECYT Regular Numbers 1150345 and 1170121.



\label{lastpage}

\end{document}